\documentclass[fleqn,twoside]{article}
\usepackage[headings]{espcrc2}

\readRCS
$Id: espcrc2.tex,v 1.2 2004/02/24 11:22:11 spepping Exp $
\usepackage{epsf}

\usepackage[figuresright]{rotating}

\newcommand{\geve}{~GeV}
\newcommand{\gevp}{~GeV/$c$}
\newcommand{\gevm}{~GeV/$c^2$}
\newcommand{\phione}{$\phi^{}_1$}
\newcommand{\phitwo}{$\phi^{}_2$}
\newcommand{\belle}{Belle}
\newcommand{\babar}{BaBar}
\newcommand{\cp}{$CP$}
\newcommand{\mbc}{$m^{}_{\rm bc}$}
\newcommand{\deltaE}{$\Delta E$}

\newcommand{\cpipi}{${\cal C}^{}_{\pi\pi}$}
\newcommand{\spipi}{${\cal S}^{}_{\pi\pi}$}
\newcommand{\crhopi}{${\cal C}^{}_{\rho\pi}$}
\newcommand{\srhopi}{${\cal S}^{}_{\rho\pi}$}
\newcommand{\crhorho}{${\cal C}^{}_{\rho\rho}$}
\newcommand{\srhorho}{${\cal S}^{}_{\rho\rho}$}

\newcommand{\bbar}{\overline{B}{}^{\,0}}
\newcommand{\bbbar}{$B^0$-$\bbar$}

\newcommand{\ra}{\!\rightarrow\!}

\newcommand{\bpipi}{$B^0\ra\pi^+\pi^-$}
\newcommand{\brhopi}{$B^0\ra\rho^+\pi^-$}
\newcommand{\brhorho}{$B^0\ra\rho^+\rho^-$}

\newcommand{\AmS}{{\protect\the\textfont2
  A\kern-.1667em\lower.5ex\hbox{M}\kern-.125emS}}

\title{
\begin{flushright}
{\large UCHEP-04-01}
\end{flushright} 
\vskip-0.05in
Constraints upon the CKM angle \phitwo\ from \belle\ and \babar }

\author{A.\ J.\ Schwartz\address{
Physics Department \\
University of Cincinnati \\
P.O. Box 210011\\
Cincinnati, Ohio 45221 }
}
       
\runtitle{Constraints upon the CKM angle \phitwo}
\runauthor{A.\ Schwartz}

\begin{document}

\begin{abstract}
The \belle\ and \babar\ experiments have measured branching 
fractions and \cp\ asymmetries in the charmless decay modes 
\bpipi,  $B^0\ra\rho^\pm\pi^\mp$, and \brhorho. From these 
measurements, contraints upon the CKM angle \phitwo\ can be 
obtained. These constraints consistently indicate that 
\phitwo\ is around~100$^\circ$.
\vspace{1pc}
\end{abstract}

\maketitle

\section{INTRODUCTION}

The Standard Model predicts \cp\ violation to occur in $B^0$
meson decays owing to a complex phase in the $3\!\times\!3$
Cabibbo-Kobayashi-Maskawa (CKM) mixing matrix. This phase
is illustrated by plotting the unitarity condition
$V^*_{ub} V^{}_{ud} + V^*_{cb} V^{}_{cd} + V^*_{tb} V^{}_{td} =0$ 
as vectors in the complex plane: the phase results in 
a triangle of nonzero height.  One interior angle of the 
triangle, denoted \phione\ or $\beta$, is determined from 
$B^0\ra J/\psi\,K^0$ decays.\footnote{Charge-conjugate modes 
are included throughout this paper unless noted otherwise.}
Another interior angle, \phitwo\ or $\alpha$, is determined 
from charmless decays such as \bpipi, \brhopi, and \brhorho.
To determine \phitwo\ requires measuring time-dependent decay 
rates; here we present such measurements from the 
\belle~\cite{belle} and \babar~\cite{babar} experiments.

\begin{figure}
\centerline{\epsfxsize=4.8cm \epsfbox{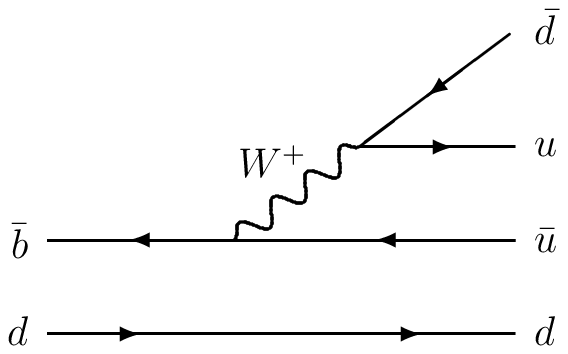}}
\vskip0.25in
\centerline{\epsfxsize=4.8cm \epsfbox{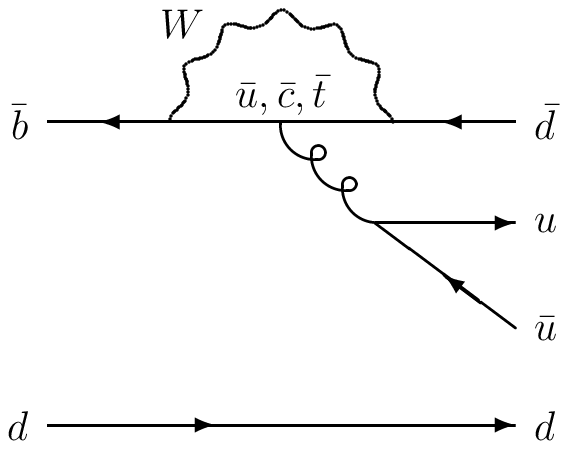}}
\vskip-0.10in
\caption{Tree-level diagram (top) and penguin diagram
(bottom) for \bpipi, \brhopi, and \brhorho\ decays.}
\label{fig:intro1}
\end{figure}

In neutral $B$ meson decays, \cp\ violation arises 
predominantly because of interference between a $B^0\ra f$ 
decay amplitude and a $B^0\ra\bbar\ra f$ amplitude. For the 
final states considered here, there are two decay amplitudes 
possible: a $b\ra u$ ``tree'' and a $b\ra d$ ``penguin'' 
(see Fig.~\ref{fig:intro1}). Because these amplitudes have 
different weak phases, additional information is needed 
to determine \phitwo, such as the size of the penguin 
amplitude or the difference in strong phases between 
the penguin and tree amplitudes. 

\section{ANALYSIS}

The analyses of \bpipi, \brhopi, and \brhorho\ decays have 
several similarities. Events are selected by requiring two 
opposite-charge pion-candidate tracks originating from the 
interaction region, and appending zero, one, or two $\pi^0$'s. 
The charged pion identification criteria are based on 
information from either a DIRC detector (\babar)~\cite{babarPID}
or time-of-flight counters and aerogel cherenkov 
counters (\belle)~\cite{bellePID}. Both experiments also 
use $dE/dx$ information from the central tracking chamber.

$B$ decays are identified via two
kinematic variables: the ``beam-constrained'' mass, \mbc, 
and the energy difference, \deltaE. The former 
is defined as $\sqrt{E^2_b-p^2_B}$ and 
the latter as $E^{}_B-E^{}_b$, where
$p^{}_B$ is the reconstructed $B$ momentum, 
$E^{}_B$ is the reconstructed $B$ energy,
and $E^{}_b$ is the beam energy, all evaluated in 
the $e^+e^-$ center-of-mass (CM) frame.
After selection cuts, the \mbc\ and \deltaE\ distributions
are jointly fit for the signal event yields. 
This fit includes contributions from backgrounds,
whose \mbc-\deltaE\ distributions are obtained from
either Monte Carlo (MC) simulation or extrapolation
from \mbc-\deltaE\ sidebands.

A tagging algorithm is used to identify the flavor of the 
$B$ signal decay, i.e., whether it is $B^0$ or~$\bbar$.
This algorithm examines tracks not associated with the 
signal decay to identify the flavor of the non-signal~$B$. 
It depends predominantly on identifying leptons or
kaons. The signal-side tracks are fit for a signal 
decay vertex, and the tag-side tracks are fit for 
a tag-side decay vertex; the distance $\Delta z$ 
between vertices is to good approximation proportional 
to the time difference between the $B$ decays: 
$\Delta z\approx (\beta\gamma c)\Delta t$, where
$\beta\gamma$ is the Lorentz boost of the $e^+e^-$
system and equals 0.43 (0.56) for \belle\ (\babar).
One subsequently does an unbinned maximum likelihood (ML)
fit to $\Delta t$ to measure or constrain~\phitwo.

The dominant background for all three decays is
$e^+e^-\ra q\bar{q}$ continuum events, where
$q=u,d,s,c$. To distinguish such events from
$e^+e^-\ra B\overline{B}$ events, the event topology 
is used: in the CM frame, continuum events tend to be 
collimated along the beam directions while $B\overline{B}$ 
events tend to be spherical. In \belle, the ``shape'' of 
an event is typically quantified via Fox-Wolfram 
moments~\cite{FoxWolfram} of the form
$h^{}_\ell = \sum_{i,j} p^{}_i\,p^{}_j\,P^{}_\ell(\cos\theta^{}_{ij})$,
where $i$ runs over all tracks on the tagging side and
$j$ runs over all tracks on either the tagging side or
the signal side. The function $P^{}_\ell$ is the $\ell$th
Legendre polynomial and $\theta^{}_{ij}$ is the angle 
between momenta $\vec{p^{}_i}$ and $\vec{p^{}_j}$.
These moments are combined into a Fisher 
discriminant~\cite{Fisher_discr}, and the discriminant
is subsequently combined with the probability
density function (pdf) for the cosine of the angle
between the $B$ direction and the electron beam
direction. This yields an overall likelihood~${\cal L}$, 
which is evaluated for both a $B\overline{B}$ hypothesis 
and a continuum hypothesis. Signal $B\ra f$ events are 
separated from continuum events by cutting on the 
likelihood ratio ${\cal L}_{B\overline{B}}/
({\cal L}_{B\overline{B}}+ {\cal L}_{q\bar{q}})$.

In \babar, $B\ra f$ signal is separated from continuum
background using several methods.
For \bpipi, a cut $|\cos\theta^{}_{\rm sph}|<0.8$ is imposed, where
$\theta^{}_{\rm sph}$ is the angle between the sphericity axis of
the $B$ candidate and that of the rest of the event. 
A Fisher discriminant (${\cal F}$) is then constructed
from $\sum_{i} p^{}_i$ and $\sum_{i} p^{}_i |\cos\theta^{}_i |^2$,
where $p^{}_i$ is the momentum of particle $i$, $\theta^{}_i$
is the angle between $\vec{p}^{}_i$ and the $B$ thrust axis
(both evaluated in the $e^+e^-$ CM frame), and $i$ runs over 
all particles not associated with the $B$ decay. A pdf for 
${\cal F}$ is included in the ML fit to~$\Delta t$. For \brhopi\ 
and one~\cite{babar_rhorho2} of two \brhorho\ analyses, a neural 
network is used that includes the two event-shape variables from 
${\cal F}$. The output of the neural network is included in the 
$\Delta t$ fit. For the other \brhorho\ analysis~\cite{babar_rhorho1},
a cut $|\cos\theta^{}_{\rm th}|<0.8$ is made, where $\theta^{}_{\rm th}$ 
is the angle between the thrust axis of the $B$ candidate and 
that of the rest of the event. The analysis subsequently uses 
a Fisher discriminant constructed from 11 observables.

\section{{\boldmath \bpipi}}

The decay time dependence of $B^0/\bbar\ra\pi^+\pi^-$ 
decays is given by~\cite{gronau}
\begin{eqnarray}
\frac{dN}{d\Delta t} & \!\!\!\propto\!\!\! & 
e^{-\Delta t/\tau}\biggl[1 - q\,{\cal C}^{}_{\pi\pi}\cos(\Delta m\Delta t)
\nonumber \\
 & & \hskip0.90in 
+ q\,{\cal S}^{}_{\pi\pi}\sin(\Delta m \Delta t)\,\biggr], 
\label{eqn:pipi}
\end{eqnarray}
where $q\!=\!+1$ ($q\!=\!-1$) corresponds to $B^0$ ($\bbar$) 
tags, and $\Delta m$ is the \bbbar\ mass difference. The 
parameters \cpipi\ and \spipi\ are \cp-violating and 
related to \phitwo\ via~\cite{gronaurosner}
\begin{eqnarray}
{\cal C}^{}_{\pi\pi} & \!\!\!=\!\!\! & \frac{1}{R}\cdot
\left( 2\left|\frac{P}{T}\right|\sin(\phi^{}_1 - \phi^{}_2)\sin\delta\right)
\label{eqn:cpipi} \\
 & & \nonumber \\
{\cal S}^{}_{\pi\pi} & \!\!\!=\!\!\! & 
\frac{1}{R}\cdot
\biggl( 2\left|\frac{P}{T}\right|\sin(\phi^{}_1-\phi^{}_2)\cos\delta +
 \nonumber \\
 & & \hskip0.50in 
 \sin\,2\phi^{}_2 - \left|\frac{P}{T}\right|^2\sin\,2\phi^{}_1 \biggr)
\label{eqn:spipi} \\
 & & \nonumber \\
R & \!\!\!=\!\!\! & 
1-2\left|\frac{P}{T}\right|\cos(\phi^{}_1+\phi^{}_2)\cos\delta + 
\left|\frac{P}{T}\right|^2\,,
\end{eqnarray}
where $\phi^{}_1=(23.2^{+1.6}_{-1.5})^\circ$~\cite{hfag},
$|P/T|$ is the magnitude of the penguin amplitude relative
to that of the tree amplitude, and $\delta$ is the strong phase 
difference between the two amplitudes. If there were no penguin 
contribution, $P=0$, ${\cal C}^{}_{\pi\pi}=0$, and 
${\cal S}^{}_{\pi\pi} = \sin 2\phi^{}_2$.
Since Eqs.~(\ref{eqn:cpipi}) and (\ref{eqn:spipi}) have three
unknown parameters, measuring \cpipi\ and \spipi\ determines 
a volume in $\phi^{}_2$ - $\delta$ - $|P/T|$ space.

\begin{figure}[t]
\centerline{\epsfxsize=7.5cm \epsfbox{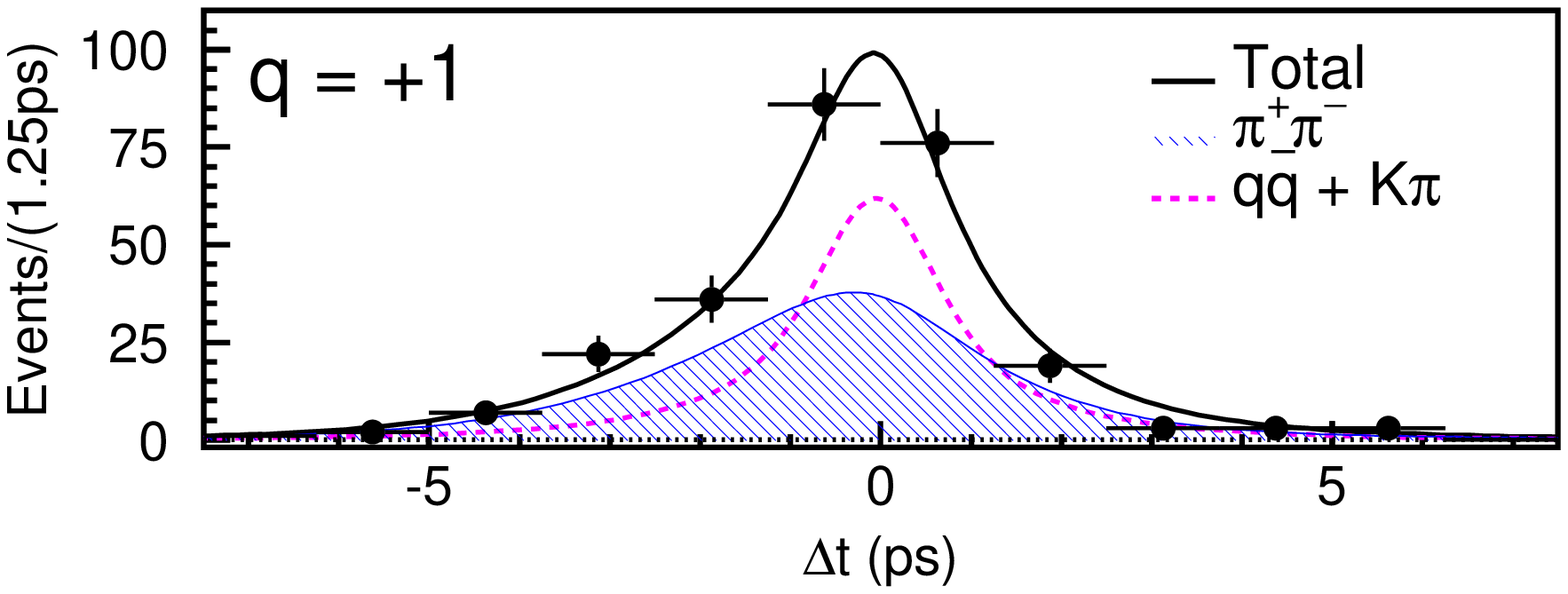}} 
\centerline{\epsfxsize=7.5cm \epsfbox{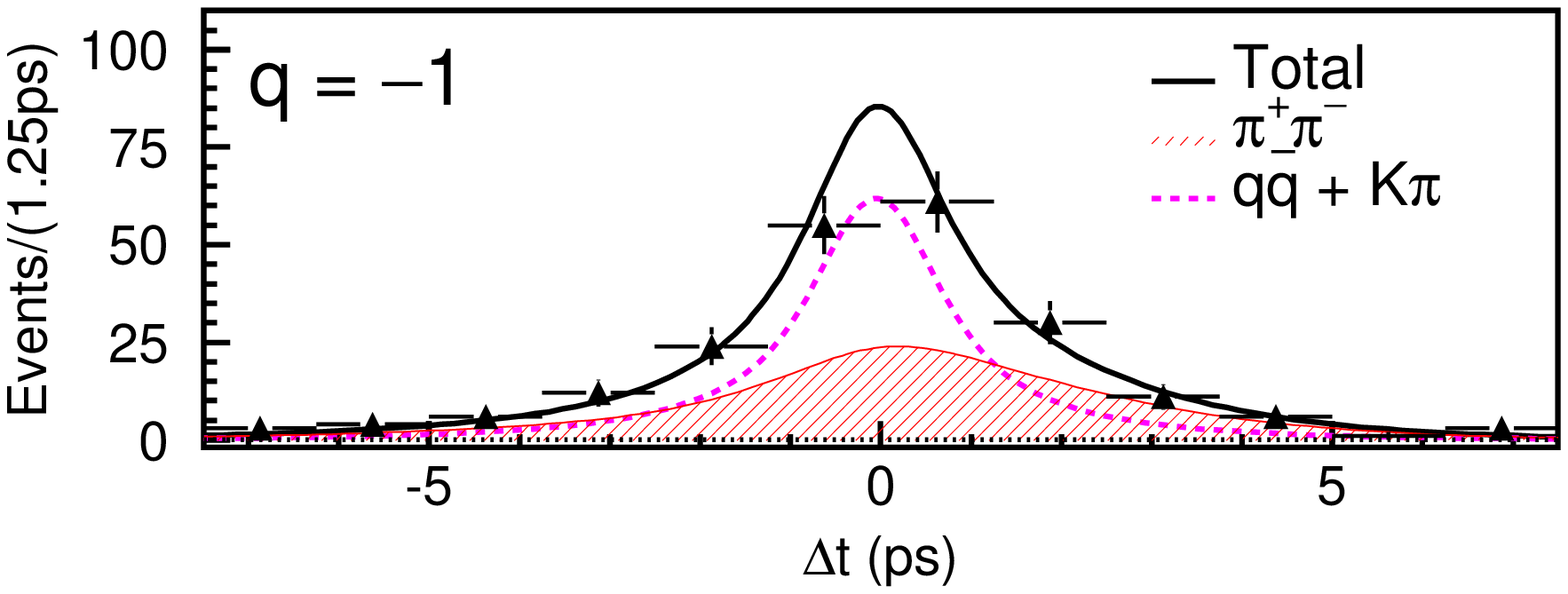}}
\centerline{\epsfxsize=7.5cm \epsfbox{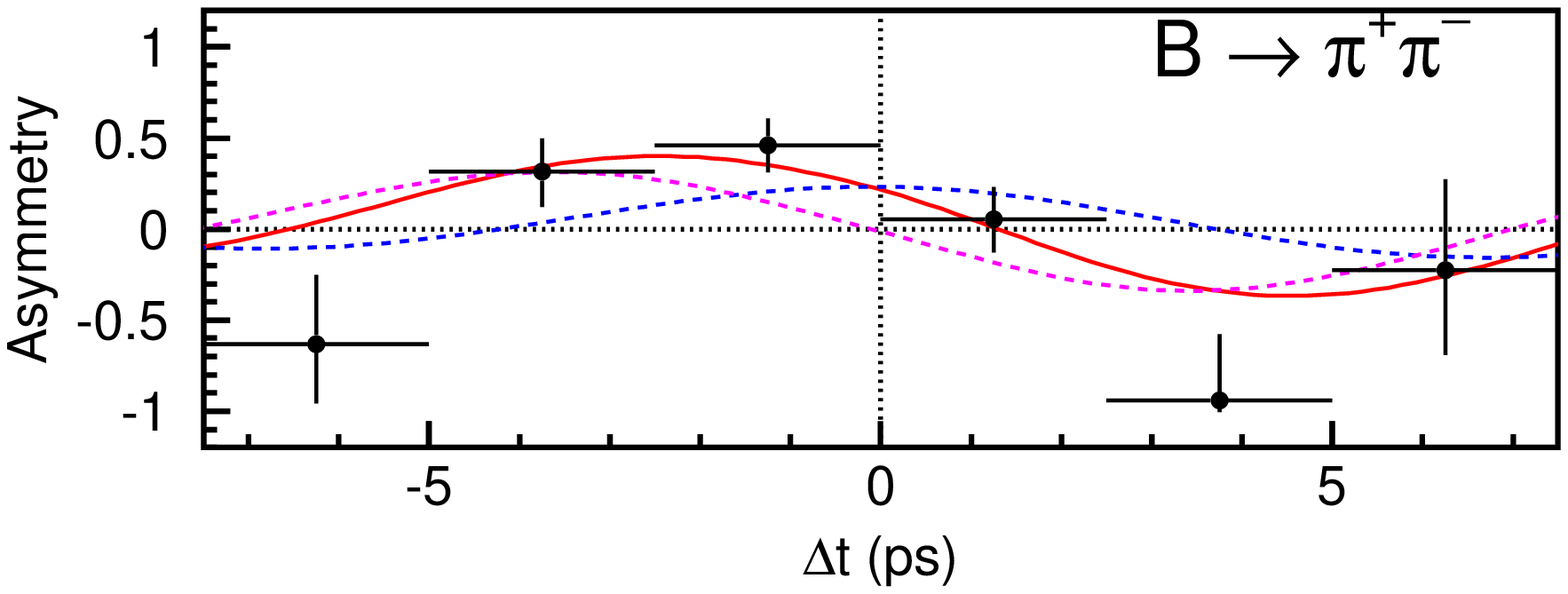}}
\vskip-0.26in
\caption{Belle results for \bpipi~\cite{belle_pipi}: 
the $\Delta t$ distributions of $q=1$ tags (top), $q=-1$ 
tags (middle), and the resulting \cp\ asymmetry (bottom). 
The smooth curves are projections of the unbinned ML fit. }
\label{fig:pipi1}
\end{figure}

The most recent \belle\ measurement of \cpipi\ and \spipi\ is 
with 140~fb$^{-1}$ of data~\cite{belle_pipi}. Candidates must 
satisfy $5.271{\rm\ GeV}/c^2\!<\!m^{}_{\rm bc}\!<\!5.287{\rm\ GeV}/c^2$
and $\Delta E\!<\!0.064$\geve; the final event sample consists 
of 224 $\bbar\ra\pi^+\pi^-$ candidates and 149 \bpipi\ 
candidates after background subtraction. The ratio of 
signal to background is $\sim$\,0.3. These events are 
subjected to an unbinned ML fit to~$\Delta t$, in which 
additional pdf's and resolution functions are included 
to account for backgrounds. There are two free parameters 
in the fit, and the results are 
${\cal C}^{}_{\pi\pi}\!=\!-0.58\,\pm\,0.15\,({\rm stat})\,\pm\,0.07\,({\rm syst})$ and
${\cal S}^{}_{\pi\pi}\!=\!-1.00\,\pm\,0.21\,({\rm stat})\,\pm\,0.07\,({\rm syst})$.
These values are consistent with previous 
\belle\ measurements~\cite{belle_pipi_previous} and
indicate large \cp\ violation. Fig.~\ref{fig:pipi1} 
shows the $\Delta t$ distributions for the $q=\pm 1$ 
samples; a clear difference is seen between the
distributions. Many cross-checks have been done 
for this analysis, including an independent ``blind'' 
analysis using a binned ML fit. The latter results 
are very close to those of the main fit.

\begin{figure}
\centerline{\epsfxsize=7.0cm \epsfbox{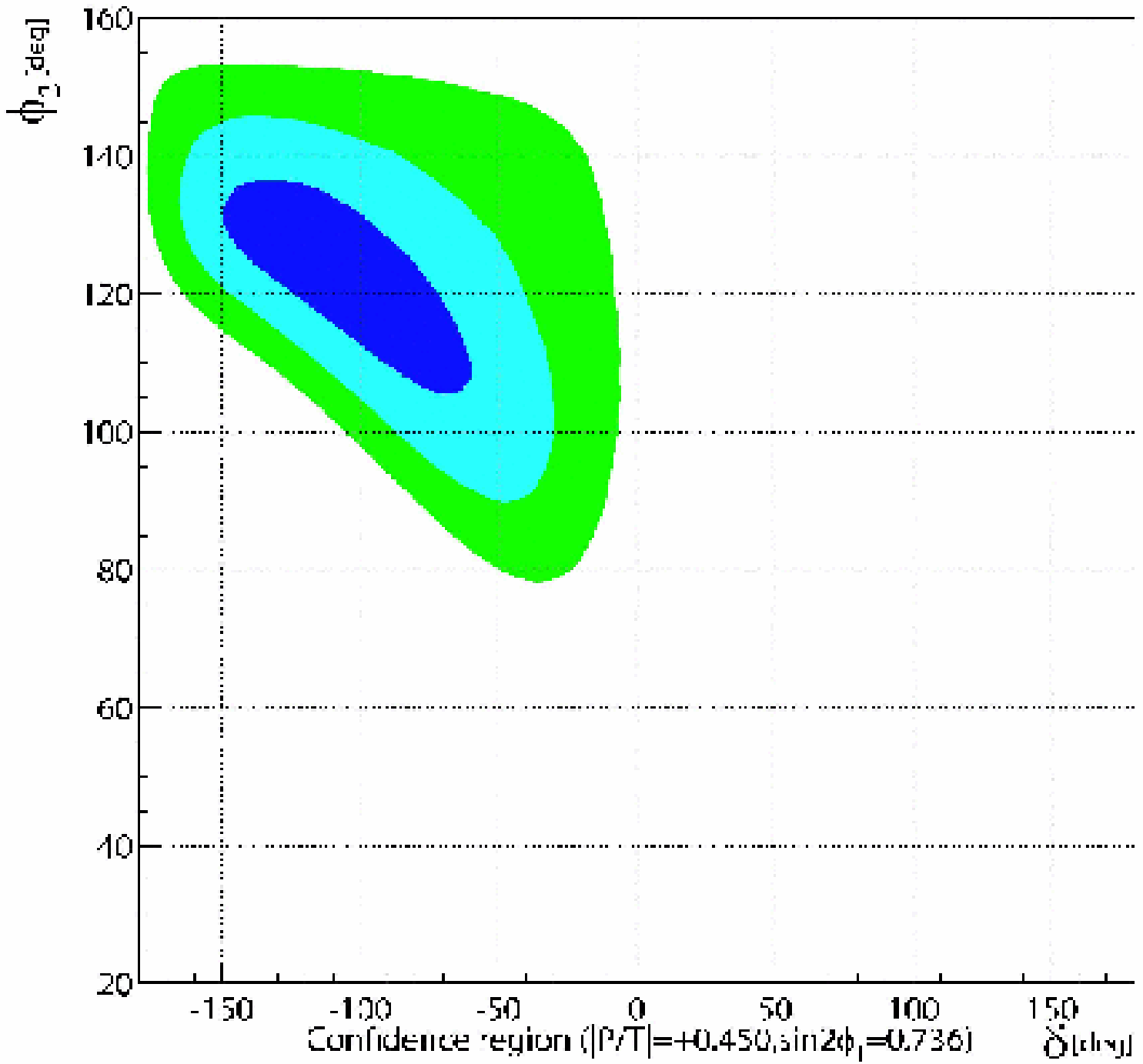}}
\vskip0.30in
\centerline{\epsfxsize=7.0cm \epsfbox{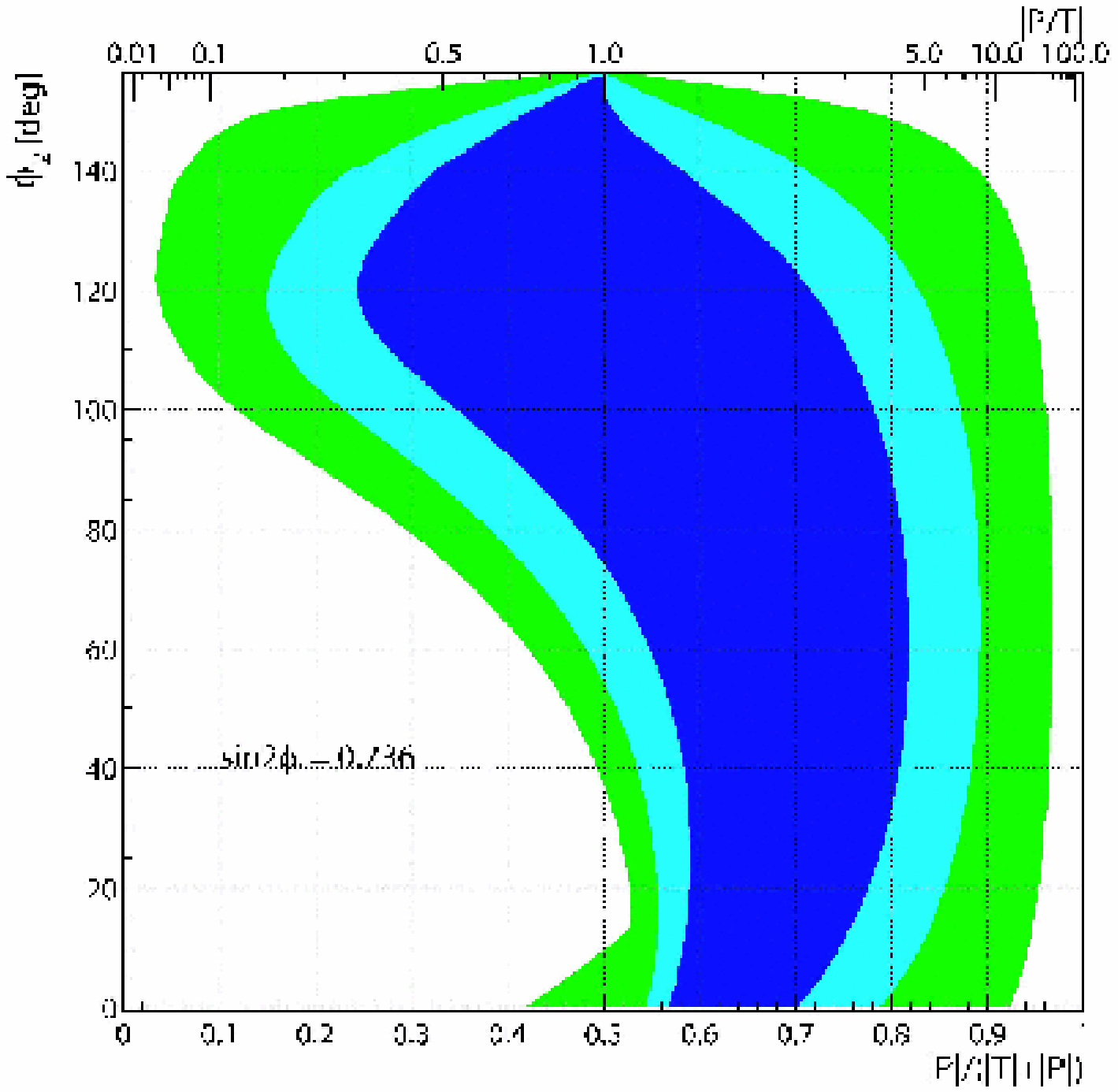}}
\vskip-0.20in
\caption{\belle\ results for \bpipi: 
{\it (a)}\ constraints in the $\phi^{}_2$-$\delta$ plane for
$|P/T|<0.45$; and 
{\it (b)}\ constraints in the $\phi^{}_2$ - $P/(|P|+|T|)$ 
plane for all values of $\delta$.
The dark blue region corresponds to $1\sigma$ C.L., 
the light blue region to 90\% C.L., and the green
region to 95\% C.L. }
\label{fig:pipi2}
\end{figure}

The \belle\ values for \cpipi\ and \spipi\ prescribe a 
95\% C.L. volume in $\phi^{}_2$ - $\delta$ - $|P/T|$ space.
Slicing this volume at fixed $|P/T|$ gives a 95\% C.L.\ constraint 
in the \phitwo-$\delta$ plane; slicing this volume at fixed
$\delta$ gives a constraint in the \phitwo-$|P/T|$ plane.
Two such projections are shown in Figs.~\ref{fig:pipi2}a 
and \ref{fig:pipi2}b; the resulting constraints are
$90^\circ < \phi^{}_2 < 146^\circ$ for $|P/T|<0.45$
(as predicted by QCD factorization~\cite{qcd_fctrztn} 
and perturbative QCD~\cite{pert_qcd}), and 
$|P/T|>0.17$ for any value of $\delta$.

The \babar\ experiment has also measured \cpipi\ 
and \spipi\ using an unbinned ML fit~\cite{babar_pipi}. 
The most recent result is from 205~fb$^{-1}$ of 
data~\cite{babar_pipi_new}; the values obtained are
${\cal C}^{}_{\pi\pi} = -0.09\,\pm\,0.15\,({\rm stat})\,\pm\,0.04\,({\rm syst})$ 
and ${\cal S}^{}_{\pi\pi}=-0.30\,\pm\,0.17\,({\rm stat})\,\pm\,0.03\,({\rm syst})$.
These values 
are inconsistent with the \belle\ result at the level 
of~$3.2\sigma$~\cite{hfag}.
The \babar\ analysis differs from that of \belle\ in that
fewer cuts are made to enrich the data sample; rather,
additional pdf's for the discriminating variables are 
included in the likelihood function.
A total of 68\,030 events are fit, and a signal yield of 
$467\pm 33$ \bpipi\ decays is obtained. There are 46 free 
parameters (including \cpipi\ and \spipi) in the fit. 

The \belle\ and \babar\ results can be averaged together 
to constrain~\phitwo. However, such a constraint requires
knowledge or assumptions about $|P/T|$ or $\delta$. 
A model based on $SU(3)$ symmetry (and including an 
$SU(3)$-breaking factor $f^{}_K/f^{}_\pi$ for tree 
amplitudes) and the measured rates 
for $B^0\ra K^0\pi^+$ and $B^0\ra K^+\pi^-$ indicates 
$\phi^{}_2\!=\!(103\,\pm\,17)^\circ$~\cite{gronaurosner2}
(note: this uses the \babar\ result for 113~fb$^{-1}$ of data).
A preferred method is to use isopin symmetry and the 
measured rates for $B^+\ra\pi^+\pi^0$, $B^0\ra\pi^0\pi^0$,
and charge-conjugates; this method can determine \phitwo\ 
with little theoretical uncertainty~\cite{gronaulondon}.
However, the decay $B^0\ra\pi^0\pi^0$ has only recently
been observed and the asymmetry between $B^0$ and $\bbar$
measured~\cite{belle_pi0pi0}. Future measurements with
higher statistics should yield an interesting constraint 
on~\phitwo. The overall $(B^0+\bbar)\ra\pi^0\pi^0$ 
branching fraction can be used to obtain an upper 
bound~\cite{glss} on the angular difference
$\theta\equiv \phi^{}_2-\phi^{}_{2\,{\rm eff}}$,
where $S^{}_{\pi\pi}=\sin 2\phi^{}_{2\,{\rm eff}}$
(i.e., $\phi^{}_{2\,{\rm eff}}\ra\phi^{}_2$ as $P\ra 0$).
Using \cpipi\ and the most recent values of the 
above branching fractions~\cite{hfag} fluctuated
by $1\sigma$ in the conservative direction, one 
obtains $\theta<36^\circ$.

\section{{\boldmath \brhopi}}

For \brhopi, the final state is not a \cp\ eigenstate. 
There are thus four separate decays to consider: 
$B^0\ra\rho^\pm\pi^\mp$ and $\bbar\ra\rho^\pm\pi^\mp$. 
The decay rates can be parametrized as~\cite{rhopi}
\begin{eqnarray}
\frac{dN(B\ra\rho^\pm\pi^\mp)}{d\Delta t} & \!\!\!\propto\!\!\! & 
(1\pm A^{\rho\pi}_{CP})\ \times  \nonumber \\
 & & \hskip-0.90in
e^{-\Delta t/\tau}\biggl[1 - q\,({\cal C}^{}_{\rho\pi}\pm 
\Delta {\cal C}^{}_{\rho\pi})\cos(\Delta m\Delta t)
\nonumber \\
 & & \hskip-0.50in 
+\ q\,({\cal S}^{}_{\rho\pi}\pm \Delta {\cal S}^{}_{\rho\pi})
\sin(\Delta m \Delta t)\,\biggr], 
\label{eqn:rhopi}
\end{eqnarray}
where $q\!=\!+1$ ($q\!=\!-1$) corresponds to $B^0$ ($\bbar$) 
tags. The parameters \crhopi\ and \srhopi\ are \cp-violating, 
while the parameters $\Delta{\cal C}^{}_{\rho\pi}$ 
and $\Delta{\cal S}^{}_{\rho\pi}$ are \cp-conserving. 
$\Delta{\cal C}^{}_{\rho\pi}$ characterizes the 
difference in rates between ``$W\ra\rho$'' processes
$B^0\ra\rho^+\pi^-$ or $\bbar\ra\rho^-\pi^+$ and 
``${\rm spectator}\ra\rho$'' processes
$B^0\ra\rho^-\pi^+$ or $\bbar\ra\rho^+\pi^-$ 
(see~Fig.~\ref{fig:intro1}).
$\Delta{\cal S}^{}_{\rho\pi}$ depends, in addition,
on differences in phases between $W\ra\rho$ 
and ${\rm spectator}\ra\rho$ amplitudes.

The parameter $A^{\rho\pi}_{CP}$ is equal to the time 
and flavor integrated asymmetry:
$\Gamma(B^0\ra\rho^+\pi^-) + \Gamma(\bbar\ra\rho^+\pi^-)  
- \Gamma(\bbar\ra\rho^-\pi^+) - \Gamma(B^0\ra\rho^-\pi^+)$
divided by the sum of the four rates. We also define
two separate \cp\ asymmetries: 
\begin{eqnarray}
A^{}_{+-} & \equiv & \frac{N(\bbar\ra\rho^-\pi^+)-N(B^0\ra\rho^+\pi^-)}
	{N(\bbar\ra\rho^-\pi^+)+N(B^0\ra\rho^+\pi^-)} \nonumber \\
 & = & -\,\frac{A^{\rho\pi}_{CP} + C^{}_{\rho\pi} + A^{\rho\pi}_{CP}\cdot\Delta C^{}_{\rho\pi}}
     {1 + \Delta C^{}_{\rho\pi} + A^{\rho\pi}_{CP}\cdot C^{}_{\rho\pi}}  \\
{\rm and} & & \nonumber \\
A^{}_{-+} & \equiv & \frac{N(\bbar\ra\rho^+\pi^-)-N(B^0\ra\rho^-\pi^+)}
	{N(\bbar\ra\rho^+\pi^-)+N(B^0\ra\rho^-\pi^+)}\nonumber \\
& = & \frac{A^{\rho\pi}_{CP} - C^{}_{\rho\pi} - A^{\rho\pi}_{CP}\cdot\Delta C^{}_{\rho\pi}}
     {1 - \Delta C^{}_{\rho\pi} - A^{\rho\pi}_{CP}\cdot C^{}_{\rho\pi}}\ .
\end{eqnarray} 
$A^{}_{+-}$ depends only on $W\ra\rho$ processes and 
$A^{}_{-+}$ depends only on ${\rm spectator}\ra\rho$ processes.

Both \babar\ and \belle\ have done unbinned ML fits to the
$\Delta t$ distributions of $B^0\ra\rho^\pm\pi^\pm$ decays 
to determine $A^{\rho\pi}_{CP},\ {\cal C}^{}_{\rho\pi},\ 
{\cal S}^{}_{\rho\pi},\ \Delta{\cal C}^{}_{\rho\pi}$,
$\Delta{\cal S}^{}_{\rho\pi}$, $A^{}_{+-}$, and $A^{}_{-+}$.
The \belle\ analysis is with 140~fb$^{-1}$ of data~\cite{belle_rhopi};
the \babar\ analysis, originally with 81~fb$^{-1}$ of data~\cite{babar_rhopi1},
has been updated with 113~fb$^{-1}$~\cite{babar_roos}.

To remove charge-ambiguous decays and possible interference
between \brhopi\ and $B^0\ra\rho^-\pi^+$ amplitudes, one must 
eliminate the overlap region of the $\pi^+\pi^-\pi^0$ Dalitz 
plot. \belle\ does this by requiring 
$0.57{\rm\ GeV}/c^2\!<\!m^{}_{\pi^\pm\pi^0}\!<\!0.97$\gevm\ 
and $m^{}_{\pi^\mp\pi^0}\!>\!1.22$\gevm. \babar\ makes the 
looser selection 
$0.40{\rm\ GeV}/c^2\!<\!m^{}_{\pi^\pm\pi^0}\!<\!1.30$\gevm\ 
and requires that $m^{}_{\pi^\mp\pi^0}$ not be in this range.
In addition, \babar\ requires that the bachelor
track from $B\ra\rho\pi$ has $p>2.4$\gevp, where $p$ is
evaluated in the $e^+e^-$ CM frame; only 14\% of pions 
from (selected) $\rho^\pm$ decays satisfy this requirement. 
Finally, \belle\ requires $m^{}_{\pi^+\pi^-}>0.97$\gevm\ 
to avoid the overlap region between $B^0\ra\rho^0\pi^0$
and $B^0\ra\rho^\pm\pi^\mp$.

\begin{figure}[t]
\centerline{\epsfxsize=8.0cm \epsfbox{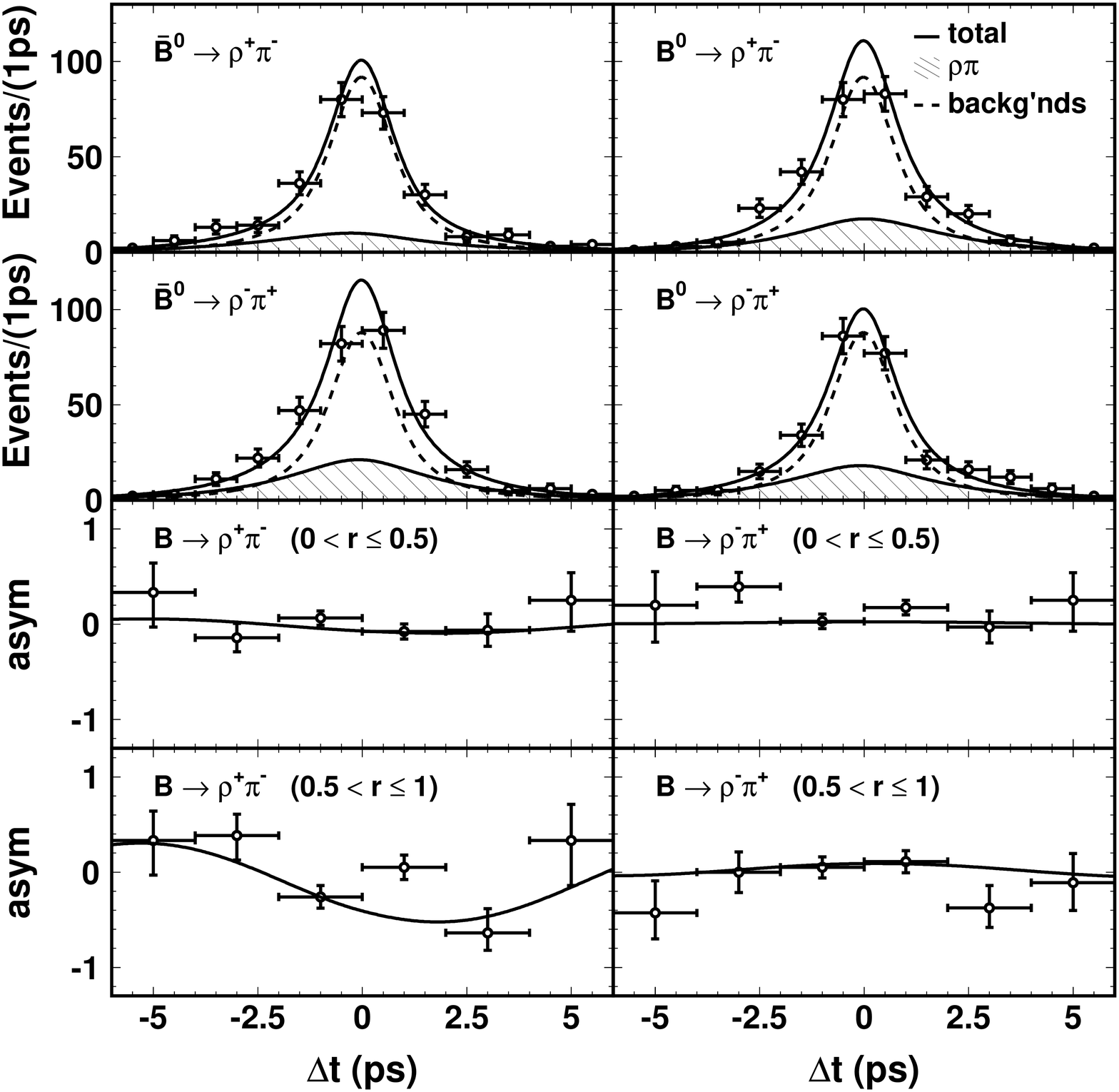}}
\vskip-0.20in
\caption{\belle\ results for \brhopi~\cite{belle_rhopi}: 
the $\Delta t$ distributions of $q=1$ tags (left), $q=-1$ 
tags (right), and the resulting \cp\ asymmetry (bottom 
rows). The asymmetry is shown separately for high-quality 
tags ($r>0.5$) and low quality tags ($r<0.5$). The 
smooth curves are projections of the unbinned ML fit.}
\label{fig:rhopi2}
\end{figure}

\belle\ subsequently defines a signal region 
$m^{}_{\rm bc}\!>\!5.27$\gevm\ and 
$-0.10~{\rm GeV}\!<\!\Delta E\!<\!0.08~{\rm GeV}$.
There are 1215 events in this region that pass all selection 
requirements. Fitting to the \mbc-\deltaE\ distributions yields
329 \brhopi\ candidates. The resulting $\Delta t$ distributions 
for $q=\pm 1$ tagged events are shown in Fig.~\ref{fig:rhopi2} 
along with projections of the unbinned ML fit in~$\Delta t$. 
Also shown is the \cp\ asymmetry, which is consistent with~zero.

\begin{figure}
\centerline{\epsfxsize=7.2cm \epsfbox{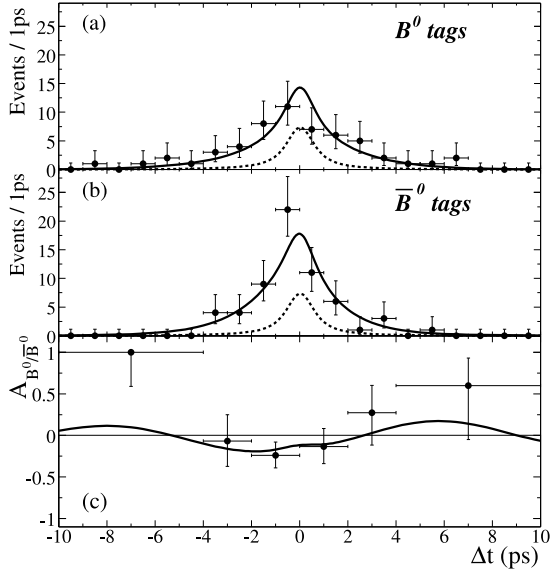}}
\vskip-0.30in
\caption{\babar\ results for \brhopi~\cite{babar_rhopi1}: 
the $\Delta t$ distributions of $q=1$ tags (top), $q=-1$ 
tags (middle), and the resulting \cp\ asymmetry (bottom). 
The smooth curves are projections of the unbinned ML fit.}
\label{fig:rhopi3}
\end{figure}

The \babar\ results are similar to those from \belle;
the corresponding $\Delta t$ distributions and \cp\ asymmetry
are shown in Fig.~\ref{fig:rhopi3}. All \belle\ and \babar\ 
results are listed in Table~\ref{tab:rhopi1}. There is very 
good agreement between the measurements except for 
$\Delta {\cal S}^{}_{\rho\pi}$, where the disagreement 
is $\sim 2\sigma$. A recent \babar\ analysis with 192~fb$^{-1}$ 
of data~\cite{babar_rhopi2} uses a different strategy than the 
quasi-two-body approach: it takes advantage of interference 
in the $\pi^+\pi^-\pi^0$ Dalitz plot as prescribed in 
Ref.~\cite{snyderquinn}. These results are also listed 
in Table~\ref{tab:rhopi1} for comparison; they are
very similar to those from the quasi-two-body analyses.

\begin{table*}[htb]
\caption{Results of fits to the $\Delta t$ 
distributions for \brhopi\ candidates.}
\label{tab:rhopi1}
\newcommand{\m}{\hphantom{$-$}}
\newcommand{\cc}[1]{\multicolumn{1}{c}{#1}}
\renewcommand{\tabcolsep}{2pc} 
\renewcommand{\arraystretch}{1.2} 
\begin{tabular}{@{}lccc}
\hline
	& {\bf Belle 2-body} & {\bf BaBar 2-body} & {\bf BaBar Dalitz} \\
 & {\bf {\boldmath (140~fb$^{-1}$)}} & {\bf {\boldmath (113~fb$^{-1}$)}} & 
    {\bf {\boldmath (192~fb$^{-1}$)}} \\
\hline
$A^{\rho\pi}_{CP}$ & $ -0.16\pm 0.10\pm 0.02$
		& $-0.114\pm 0.062\pm 0.027$ 
                & $-0.088\pm 0.049\pm 0.013$ \\
\hline
$C^{}_{\rho\pi}$ & $ 0.25\pm 0.17\,^{+0.02}_{-0.06}$
		& $0.35\pm 0.13\pm 0.05$ 
		& $0.34\pm 0.11\pm 0.05$ \\
$S^{}_{\rho\pi}$ & $-0.28\pm 0.23\,^{+0.10}_{-0.08}$
		& $-0.13\pm 0.18\pm 0.04$ 
		& $-0.10\pm 0.14\pm 0.04$ \\
\hline
$\Delta C^{}_{\rho\pi}$ & $ 0.38\pm 0.18\,^{+0.02}_{-0.04}$
		& $ 0.20\pm 0.13\pm 0.05$ 
		& $ 0.15\pm 0.11\pm 0.03$ \\
$\Delta S^{}_{\rho\pi}$ & $-0.30\pm 0.24\pm 0.09$
		& $ 0.33\pm 0.18\pm 0.03$ 
		& $ 0.22\pm 0.15\pm 0.03$ \\
\hline
$A^{}_{+-}$ & $-0.02\pm 0.16\,^{+0.05}_{-0.02}$
		& $-0.18\,\pm 0.13\,\pm\,0.05$ 
                & $-0.21\pm 0.11\,\pm\,0.04$ \\
$A^{}_{-+}$ & $-0.53\pm 0.29\,^{+0.09}_{-0.04}$
		& $-0.52\,^{+0.17}_{-0.19}\,\pm\,0.07$ 
                & $-0.47\,^{+0.14}_{-0.15}\,\pm\,0.06$ \\
\hline
\end{tabular}
\end{table*}

These measured values can be used to constrain \phitwo; 
however, since the penguin contribution is unknown, additional 
information is needed. A recent theoretical model~\cite{gronauzupan}
uses $SU(3)$ symmetry and the measured rates or limits
for branching fractions of
$B^0\ra K^{*\pm}\pi^\mp$, 
$B^0\ra\rho^\mp K^\pm$, 
$B^\pm\ra K^{*0}\pi^\pm$, and
$B^\pm\ra\rho^\pm K^0$. 
$SU(3)$-breaking effects are considered at tree level and 
accounted for via a factor $f^{}_\pi/f^{}_K$. The strong phase 
difference between the two tree amplitudes ($W\ra\rho$ and
spectator\,$\ra\rho$) is assumed to be small, as predicted
by factorization. The resulting central values and errors
for \phitwo\ are:
$102\pm 19^\circ$ for \belle\ values of \crhopi, \srhopi, 
$\Delta$\crhopi, $\Delta$\srhopi;
$93\pm 17^\circ$ for \babar\ values (113~fb$^{-1}$); and 
$95\pm 16^\circ$ for \belle\ and \babar\ values combined.

The \babar\ Dalitz plot analysis (192~fb$^{-1}$)~\cite{babar_rhopi2} 
allows one to directly fit for \phitwo\ with little 
theoretical uncertainty from the penguin contribution.
The result is
$\phi^{}_2 = (113\,^{+27}_{-17}\,({\rm stat})\,\pm\,6\,({\rm syst}))^\circ$, 
consistent with the $SU(3)$-based results above.

\section{{\boldmath \brhorho}}

The decay \brhorho\ has two vector particles 
in the final state. If the $\rho$ mesons 
are longitudinally polarized, $\ell$ is even 
and $CP=+1$; but if they are transversely polarized, 
$\ell$ can be even or odd and the final state is not 
a \cp\ eigenstate. 

For longitudinal 
polarization, \phitwo\ can be determined
from the $\Delta t$ distribution as done for \bpipi. 
However, \brhorho\ has an advantage: 
the penguin contribution is expected to be small 
relative to the tree contribution~\cite{small_penguin}, 
which reduces theoretical uncertainty on \phitwo.
Unfortunately \brhorho\ is more challenging experimentally: 
there are several backgrounds and also possible nonresonant 
contributions. The method depends upon the $\rho$'s being 
longitudinally polarized; otherwise a more involved angular 
analysis is necessary to determine~\phitwo~\cite{rhorho_ang_analysis}.
Finally, the nonnegligible decay width of the $\rho$ allows for 
$I=1$ final states, which complicates extracting \phitwo\ via 
an isospin analysis~\cite{falk}.

The decay \brhorho\ has been observed by \babar\ and the 
\cp-violating parameters \crhorho\ and \srhorho\ measured
with 81~fb$^{-1}$ of data~\cite{babar_rhorho2,babar_rhorho1}
and updated with 113~fb$^{-1}$ of data~\cite{babar_roos}. 
A similar analysis is underway at \belle\ with 250~fb$^{-1}$ of data. 
The final state consists of four pions, two charged and two neutral.
In the case of multiple \brhorho\ candidates arising from multiple
$\pi^0$ candidates, the candidate that minimizes 
the sum $\sum^{}_i (m^{(i)}_{\gamma\gamma}-m^{}_{\pi^0})$
is chosen, where $i$ runs over the $\rho^\pm$ candidates.
From MC simulation, it is found that one or more pions from 
\brhorho\ are swapped with pions from the tag side 39\% (16\%) 
of the time for longitudinal (transverse) polarization.

\babar\ selects events with relatively loose cuts and 
does an unbinned ML fit to the $\Delta t$ distribution,
including pdf's to account for backgrounds. Nonresonant 
contributions and interference with decays yielding the 
same final state, 
e.g., $B^0\ra a^{}_1\pi^0$, are estimated to be small
and neglected. For 81~fb$^{-1}$ of data, 24\,288 events 
are fit and a signal yield of $224\pm 29$ is obtained. 
The fit includes a pdf for the angles $\theta^{}_1$ and 
$\theta^{}_2$, where $\theta^{}_i$ is the angle between 
the $\pi^0$ from $\rho^\pm_i\ra\pi^\pm\pi^0$ and the $B^0$ 
in the $\rho^\pm_i$ rest frame ($i\!=\!1,2$). This pdf has 
the form~\cite{rhorho_ang_dist}
\begin{eqnarray}
\frac{d^{\,2}\Gamma}{\Gamma\,d\cos\theta^{}_1\,d\cos\theta^{}_2} & \!\!=\!\! & 
\frac{9}{4}\Bigl\{ f^{}_L\cos^2\theta^{}_1\,\cos^2\theta^{}_2\ +\ \nonumber \\
 &  & \hskip-0.20in
\left(\frac{1-f^{}_L}{4}\right)
\sin^2\theta^{}_1\,\sin^2\theta^{}_2 \Bigr\}
\end{eqnarray}
and determines $f^{}_L$, the fraction
of longitudinally polarized decays.
The fit results are 
${\cal C}^{}_{\rho\rho} = -0.23 \pm 0.24 \pm 0.14$ and
${\cal S}^{}_{\rho\rho} = -0.19 \pm 0.33 \pm 0.11$
(113~fb$^{-1}$), 
and $f^{}_L = 0.99\,\pm\,0.03\,^{+0.04}_{-0.03}$
(81~fb$^{-1}$). 
The first error listed is statistical and the second systematic.

\begin{figure}
\centerline{\epsfxsize=7.4cm \epsfbox{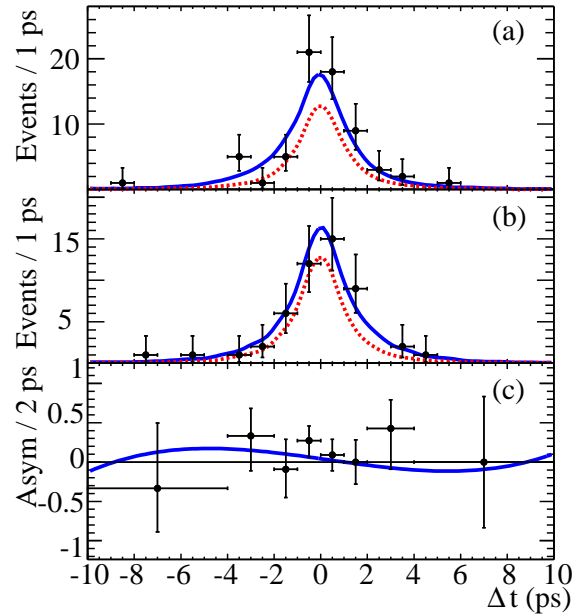}}
\vskip-0.15in
\caption{\babar\ results for \brhorho~\cite{babar_rhorho2}:
the $\Delta t$ distributions of $q=1$ tags (top), $q=-1$ 
tags (middle), and the resulting \cp\ asymmetry (bottom). 
The smooth curves are projections of the unbinned ML fit.
}
\label{fig:rhorho1}
\end{figure}

It is fortunate that $f^{}_L$ is close to unity;
in this case the final state has $CP=+1$ and
an angular analysis to determine \phitwo\ is 
unnecessary. Fig.~\ref{fig:rhorho1} shows
the $\Delta t$ distributions for $q=\pm1$ 
tagged events along with the resulting \cp\ 
asymmetry. No \cp\ violation is observed. Inputting
the measured values for \crhorho\ and \srhorho\ into 
an isospin analysis that includes the branching 
fractions for \brhorho~\cite{babar_rhorho2} and 
$B^+\ra\rho^+\rho^0$~\cite{btorho+rho0}, and the upper 
limit for $B(B^0\ra\rho^0\rho^0)$~\cite{btorho0rho0}, one
obtains $\phi^{}_2=
(96\,\pm\,10\,({\rm stat})\,\pm\,4\,({\rm syst})\,\pm\,11^{}_{\rm theory})^\circ$~\cite{giorgi_ICHEP}. 
The last error is due to the penguin contribution; it is significantly 
smaller than that for $B\ra\pi\pi$ ($\pm 36^\circ$), as expected.

\section{SUMMARY}

Time-dependent \cp\ asymmetries in \bpipi, \brhopi, and \brhorho\ 
decays are measured and used to constrain the CKM angle~\phitwo.
The \bpipi\ mode is experimentally clean but has the largest penguin 
contribution, which contributes theoretical uncertainty to~\phitwo. 
A model-independent constraint is
$90^\circ < \phi^{}_2 < 146^\circ$ for $|P/T|<0.45$ (95\% C.L.). 
An $SU(3)$-based model~\cite{gronaurosner2} indicates 
$\phi^{}_2=(103\,\pm\,17)^\circ$ (and also that $|P/T|$ 
is large). \belle\ observes large \cp\ violation in this 
mode while \babar\ does not. 

The $B^0\ra\rho^\pm\pi^\mp$ mode is more complicated as there
are more backgrounds than for \bpipi\ and the final state is 
not a \cp\ eigenstate. A model based upon $SU(3)$ symmetry 
and using the measured branching fractions for 
$B\ra K^*\pi^\pm$ and $B\ra\rho^\pm K$ obtains
$\phi^{}_2=95\pm 16^\circ$ 
(\belle\,+\,\babar\ quasi-two-body results 
combined).

The \brhorho\ mode has the smallest penguin contribution
but suffers from additional backgrounds, possible nonresonant 
contributions, and a possible $I=1$ component in the final state. 
Neglecting the latter two effects, \babar\ measures \crhorho\ 
and \srhorho\ for longitudinal polarization, which dominates 
the decay. Combining the measured values with the branching 
fractions or limits for \brhorho~\cite{babar_rhorho2}, 
$B^+\ra\rho^+\rho^0$~\cite{btorho+rho0}, and 
$B^0\ra\rho^0\rho^0$~\cite{btorho0rho0} gives 
$\phi^{}_2=
(96\,\pm\,10\,({\rm stat})\,\pm\,4\,({\rm syst})\,\pm\,11^{}_{\rm theory})^\circ$~\cite{giorgi_ICHEP}. 
This value is similar to those obtained from
measurements of $B^0\ra\pi^+\pi^-$ and 
$B^0\ra\rho^\pm\pi^\mp$ decays.

\end{document}